\documentclass[aip, apl, reprint]{revtex4-1}

\usepackage{graphicx} 
\usepackage{dcolumn} 
\usepackage{amsmath,amsfonts}
\usepackage{bm}
\usepackage{color}
\usepackage{float}
\usepackage{graphics}
\usepackage{subfigure}
\usepackage{verbatim}

\usepackage[T1]{fontenc}
\usepackage[latin9]{inputenc}
\usepackage{color}
\usepackage{units}
\usepackage{amstext}
\usepackage{graphicx}
\usepackage{amssymb}
\usepackage{fancyhdr}
\usepackage{lipsum}
\begin{document}

\title{A camera for single pixel acoustic compressive sensing in air}

\author{Jeffrey S. Rogers}
\affiliation{U.S. Naval Research Laboratory, Washington, DC 20375, USA, Code 7160}
\author{Charles A. Rohde}%
\author{Matthew D. Guild}%
\affiliation{U.S. Naval Research Laboratory, Washington, DC 20375, USA, Code 7160}
\author{Christina J. Naify}%
\affiliation{NASA Jet Propulsion Laboratory, Pasadena, CA, 91109, USA }
\author{Theodore P. Martin}%
\author{Gregory J. Orris}
\affiliation{U.S. Naval Research Laboratory, Washington, DC 20375, USA, Code 7160}
\lhead{}
\chead{}
\rhead{A camera for single pixel acoustic compressive sensing in air}
\date{\today}
\begin{abstract}
Acoustic imaging typically relies on large sensor arrays that can be electronically complex and often have large data storage requirements to process element level data. 
Recently, the concept of a \textit{single-pixel-imager} has garnered interest in the electromagnetics literature due to it's ability to form high quality images with a single receiver paired with shaped aperture screens that allow for the collection of spatially orthogonal measurements. 
Here, we present a method for creating an acoustic analog to the single-pixel-imager found in electromagnetics. 
Additionally, diffraction is considered to account for screen openings comparable to the acoustic wavelength. 
A diffraction model is presented and incorporated into the single pixel framework. 
The method is experimentally validated with laboratory measurements made in air. 
\end{abstract}
\maketitle\clearpage

\begin{figure}[b]
	\includegraphics[width=\columnwidth]{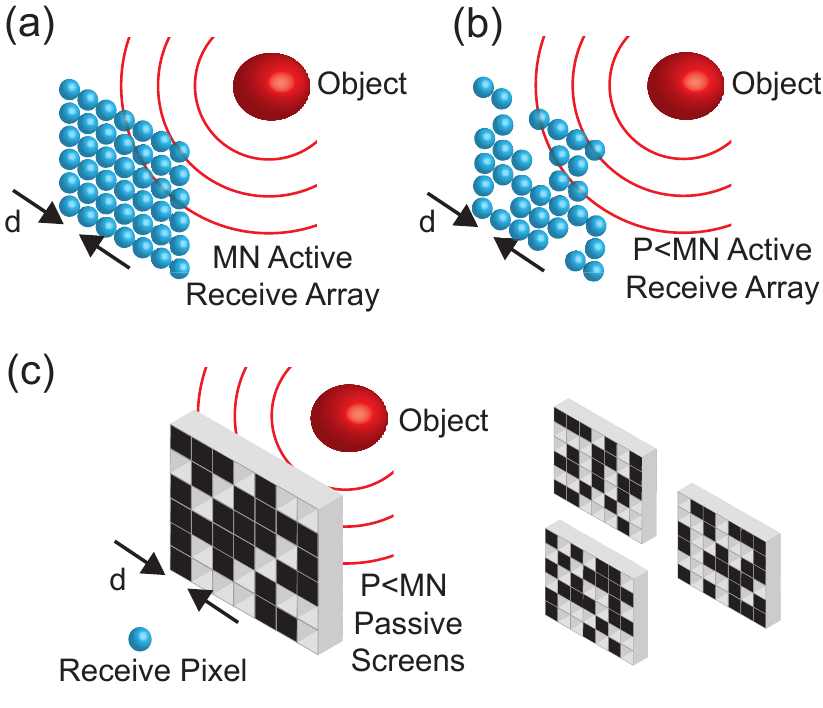}
	\caption{
		Schematic illustrating different acoustic imaging methods and their respective array geometry. 
		(a) depicts a $M\times N$  element array with sensor separation $d$ imaging an object, shown in red, emitting an acoustic signal. 
		(b) depicts a sparse sensing array used in compressive imaging where the array has been reduced to $P$ elements and $P < MN$. 
		(c) depicts the geometry assumed in single pixel imaging where there is a single receive element, denoted by the blue dot, and $P$ passive screens situated between the acoustic source and receiver.
	}
	\label{fig1}
\end{figure}

Acoustic imaging, widely used in sonar and biomedical applications, has been studied extensively in order to improve image quality with sub-diffraction resolution and reducing transducer requirements.\cite{zhang2009,zhu2011,garcia2014} 
This type of imaging, depicted in Fig.~\ref{fig1}(a), is commonly achieved using a phased detector array, and is also referred to as an acoustic camera.\cite{Brandes2007} 
While effective at target localization, the large number of required detectors and point-by-point imaging technique historically used leads to high monetary investment, electrical complexity, and collection of redundant spatial information. 
Alternatively, other imaging systems require fewer sensors but rely on mechanical scanning of the acoustic field.\cite{zhang2009} 
These scanning methods have the ability to perform comparably to fully populated phased detector arrays; but can be quite time consuming in acquiring data.
 Additionally, they require that the acoustic field statistics remain spatially stationary over the scanning time.

Compressive sensing has recently become an attractive approach due to the ability to reduce the number of measurements taken while maintaining image fidelity. \cite{candes2008}
The rationale behind compressive sensing is that most natural scenes are spatially sparse, and it is unnecessary to scan the entire scene to localize sources since most information received is redundant. 
Standard (non-compressive) imaging of a scene with $M\times N$ pixels requires taking $O(MN)$ measurements in order to image the entire scene.\cite{Qaisar2013} 
By using compressive sensing, $P$ measurements may be taken, where $P$ is less than $MN$, without any loss of image resolution as depicted in Fig.~\ref{fig1}(b). 
In addition to a smaller measurement set, the final image is inherently compressed.
This limits the amount of data which must be stored, transmitted, and analyzed. 
For a signal that is $k$-sparse, meaning there are $k \ll MN$ non-zero source locations being estimated, compressive sensing theory states that the signal can be recovered with $P$ orthogonal measurement modes where $k < P < MN$.

Although compressive sensing offers a significant advantage over traditional beamforming due to the reduced file size and fewer required measurements, a new approach called \emph{single pixel compressive imaging} has presented an even further measurement advantage. 
Single pixel imaging, depicted in Fig.~\ref{fig1}(c), presents an approach in which a single, omnidirectional detector occulted by a spatially orthogonal passive aperture to perform compressive sensing.\cite{Sun2015,Hunt2013,Duarte2011,Marcia2009,Wagadarikar2008} 
In the most straightforward implementations of this approach, originally demonstrated  in the terahertz electromagnetic frequency range, binary opaque and transparent masks are patterned to form orthogonal basis sets through which the spatial sampling is generated.\cite{Chan2008,Liutkus2014,Willett2011} 
Beampatterns are illuminated through the masks used to independently scan the scene.  
In recent years, several additional approaches, including metamaterial-based, frequency-dependent scanning have been demonstrated for electromagnetic terahertz waves.\cite{Hunt2013} 
Despite the range of approaches and results for electromagnetic imaging, very few studies have attempted single-pixel imaging in the acoustic regime. 
In 2015, Xie et al. demonstrated a single-microphone listening system to discriminate individual sound sources and their locations by exploiting frequency-diverse structures,\cite{Xie2015} while in 2016 a similar technique\cite{Gao2016} was developed for sound waves in water.

It is assumed in the electromagnetic literature  that screen openings are small relative to a wavelength and diffraction effects can therefore be neglected. 
In this manuscript we present a realization of a single-pixel acoustic camera which utilizes a compressive sensing technique coupled to a series of spatially orthogonal acoustic screens.
 To account for larger apertures, near field diffraction effects are taken considered. 
A diffraction model is derived for a 2-D geometry and formulated in the single pixel compressive sensing framework. 
The model and its resulting inversion are verified on data collected in an air acoustic experiment.

\begin{figure}[tb]
\begin{center}
\begin{tabular}{c}
\includegraphics[width=\columnwidth]{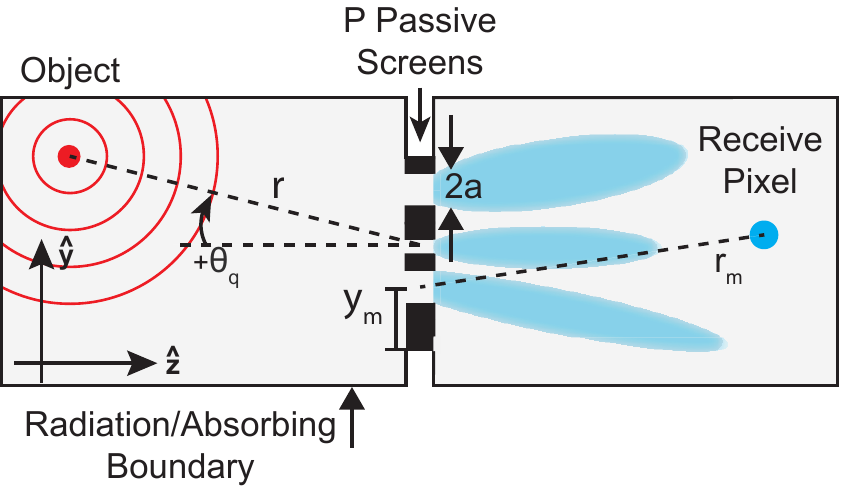}\\
\includegraphics[width=\columnwidth]{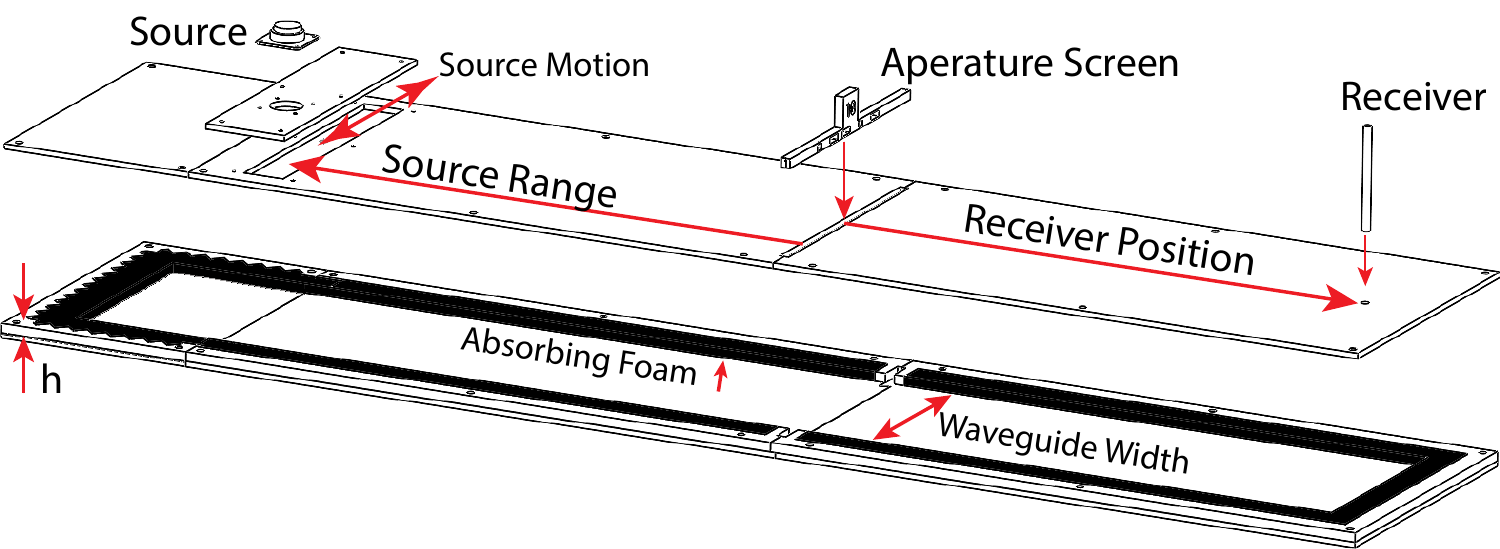}
\end{tabular}
\caption{(top) 2D experimental geometry for the acoustic single-pixel-imager. 
	An object emits an acoustic signal (red) into the waveguide with a position, $\theta$, relative to the screen's broadside. 
	Each screen has at least one opening with a minimum width, $2a$. 
	A single omnidirectional receiver (blue) is depicted on the opposite side of the screen. 
	(bottom) Experimental waveguide as built with height, $h$, and width $w$.}
\label{fig2}
\end{center}
\end{figure}

For simplicity but without loss of generality, we consider the two dimensional waveguide depicted in Fig.~\ref{fig2}.  
There is an acoustic radiating source on one side of the waveguide and a single omnidirectional receiver element on the opposite side. 
In between the acoustic source and receiver is an aperture that allows for $P$ different masks to be inserted.

The signal model used for the geometry described in Fig.~\ref{fig2} assumes a one dimensional mask and acoustic localization in azimuth given by the angle, $\theta$.  
Localization is performed at a single narrowband frequency. 
The received data, $\mathbf{y}_{P\times1}$ is a frequency domain vector given by
\begin{equation}
\label{eq01}
\mathbf{y} = \mathbf{Ax}
\end{equation} 
where the sensing matrix $\mathbf{A}=\mathbf{\Psi}\mathbf{\Phi}$ is the product of the matrix, $\mathbf{\Phi}_{M\times Q}$ (responsible for transforming the signal $\mathbf{x_{Q\times1}}$ into the spatial frequency domain) and the the matrix $\mathbf{\Psi}_{P\times M}$ (representing the measurement process in which $P$ orthogonal masks are chosen). 

In this case, the sensing matrix $\mathbf{A}$, is assumed to be known \textit{a priori}, and can be precomputed. 
For the geometry shown in Fig.~\ref{fig2}, to image the sound source in azimuth, the matrix $\mathbf{\Phi}$ would be comprised of the set of replica vectors representing the phase angle $\bm{\phi}_{m}$ of the radiated acoustic pressure $P_{m}$ from each $m^{\mathrm{th}}$ unit cell in the mask given by
\begin{equation} \label{eq02}
\tan \bm{\phi}_{m} = \frac{ \mathrm{Im} \! \left( P_{m} \right) }{ \mathrm{Re} \! \left( P_{m} \right) },
\end{equation}
The equation for the pressure field in 2D space due to a rectangular aperture of width $2a$ is given by \cite{guild2016}
\begin{align}   
	P_{m}(y,z) &= \, \frac{1}{2} p_{0} \, e^{-j \left\{ \phi_{0}^{(m)} + \frac{1}{2} k_0 \bar{\zeta}_{m} \left[1 + \left(z/\bar{\zeta}_{m} \right)^{2} \right]  \right\} } \notag \\
	 \times &\left\{ \! \mathrm{erf} \! \left[ \sqrt{ j \frac{z_{\mathrm{R}}}{\bar{\zeta}_{m}} } \left( 1 \!+\! \frac{\bar{y}_{m}}{a} \right) \right] + \mathrm{erf} \! \left[ \sqrt{ j \frac{z_{\mathrm{R}}}{\bar{\zeta}_{m}} } \left( 1 \!-\! \frac{\bar{y}_{m}}{a} \right) \right] \right\}, \label{eq03}
\end{align}
\noindent where $p_{0}$ is the source pressure, $\phi_{0}^{(m)}$ denotes the phase shift for the $m^{\mathrm{th}}$ unit cell, $k_0$ is the wavenumber in the host fluid, $z_{\mathrm{R}} = k_0a^{2}/2$ is the Rayleigh distance, erf denotes the error function, and 
\begin{align}  \label{eq04to06}
	\bar{\zeta}_{m} &= \sqrt{\bar{y}_{m}^{2} + z^{2}}, \\
	\bar{y}_{m} &= y - y_{m} - r_{m} \sin \theta_{q}, \\
	r_{m} &= \sqrt{ (y - y_{m})^{2} + z^{2}}.
\end{align}
Here, $y_{m}$ is the center of the $m^{\mathrm{th}}$ aperture and $\theta_{q} \in [-90^{\circ},90^{\circ}]$ is the azimuthal direction discretized into $Q$ angles.
The propagation delay from the sound source to each unit cell in the mask given by
\begin{equation} \label{eq07}
\bm{\phi}_{0}^{(m)}(\theta_{q}) = k_0 \sqrt{r^2+y_{m}^{2} - 2y_{m} r \sin(\theta_{q})}
\end{equation}
where $r$ is the distance from the acoustic source to the center of the mask.

\begin{figure}
\begin{center}
\includegraphics{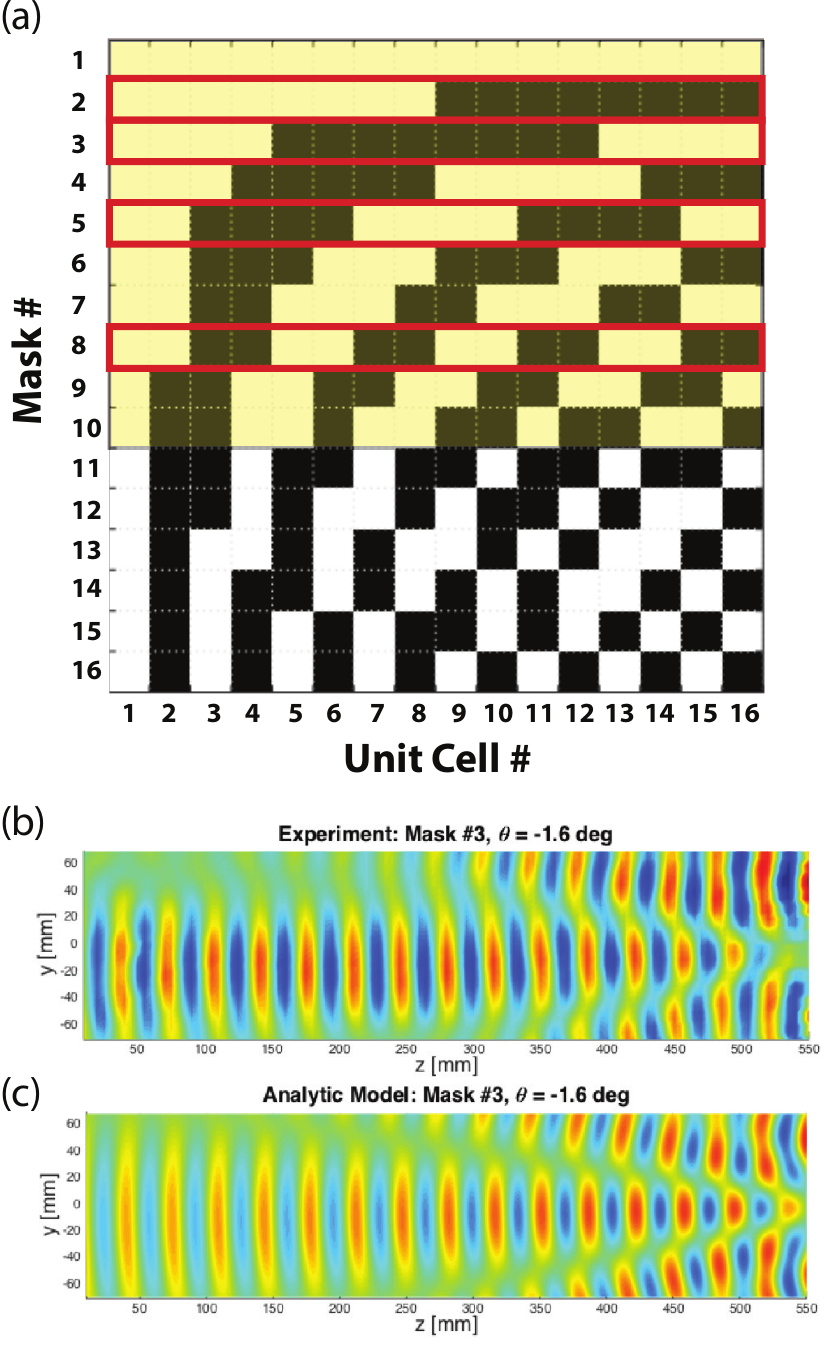}
\caption{a) Possible screen geometries forming $\mathbf{\Psi}$ for a 16 unit cell basis. Each row is a 1-D screen, used in the geometry described in Fig.~\ref{fig2}. Black squares block sound propagation and white squares are open. 
The first ten masks, highlighted in yellow, were chosen for the experimental measurements.
The masks outlined in red were used to perform the inversion and generate the results shown in Fig. \ref{fig4}.
b) Full field experimental measurement showing diffraction pattern from mask \#3 for a far-field source located at $-2^{\circ}$.
c) Full field diffraction pattern computed using the analytical model, Eqs.~(\ref{eq03})--(\ref{eq08}), for mask \#3 and a source located at $-2^{\circ}$. }
\label{fig3}
\end{center}
\end{figure}

Diffraction around the unit cells in the mask is accounted for by modeling the total pressure field radiating from each rectangular opening.  This total acoustic pressure field is illustrated in Fig.~\ref{fig3}. and can be calculated by summing the contributions from each opening in the mask by
\begin{equation}   \label{eq08}
	P_{\mathrm{tot}}(y,z,t) = e^{j \omega t} \sum_{m=1}^{M} P_{m}(y,z).
\end{equation}


A severely underdetermined set of equations must be solved to obtain $\mathbf{x}$ since it is assumed that $P \ll Q$. However, if one can assume that the data is $k$-sparse (meaning that $\mathbf{x}$ has $k$ nonzero elements) and $k < P$ the solution to Eq. \eqref{eq01} can be obtained with a $\ell_1$ minimization \cite{candes2006}
\begin{equation}
\label{eq09}
\hat{\mathbf{x}} = \min{||\mathbf{x}||_1} \hspace{.5in} \text{subject to} \hspace{.2in} \mathbf{Ax = y}
\end{equation}
where $||\mathbf{x}||_1$ denotes the $\ell_1$ norm ($\sum_{i}|A x_i - y_i|$) of $\mathbf{x}$ and $\hat{\mathbf{x}}$ is the solution of the unknown $\mathbf{x}$.

In order to guarantee exact recovery of $\mathbf{x}$, the sensing matrix $\mathbf{A}$ must satisfy the restricted isometry property (RIP) which can be expressed as \cite{candes2005} 
\begin{equation}
\label{eq10}
(1-\delta_k)||\mathbf{x}||_{\ell_2}^2 \leq ||\mathbf{Ax}||_{\ell_2}^2 \leq (1+\delta_k)||\mathbf{x}||_{\ell_2}^2
\end{equation}
where $\delta_k$ is the restricted isometry constant. 
The sensing matrix $\mathbf{A}$ satisfies the RIP for values of $\delta_k$ not too close to 1. 
In order to achieve this condition with a compressed sinusoidal basis given by $\mathbf{\Phi}$, $\mathbf{\Psi}$ can be a matrix of transmission "spikes" or 0's and 1's. 
Here, $\mathbf{\Psi}$ is chosen to be a set of orthogonal "spike combs" given by a 2-Bit quantized Discrete Cosine Transform (DCT) where the kernel is given by
\begin{equation}
\label{eq11}
\psi(p)_m = \left\{
        \begin{array}{ll}
            0 & \quad \cos \left((p-1)\pi m\right) < 0 \\
            1 & \quad \cos \left((p-1)\pi m\right) \geq 0
        \end{array}
    \right.
\end{equation}
for $p = 1,2,\ldots,P$ masks and $m = 1,2,\ldots,M$ unit cells. 
Fig.~\ref{fig3} (a) depicts the DCT kernel where each row represents one of 16 possible masks from \eqref{eq11} and each column is a unit cell. 
If the unit cell is black it represents a 0 or no transmission through the unit cell and a white unit cell is a 1 or perfect transmission through the unit cell.

It has recently been shown that computing the RIP directly for a matrix, $\mathbf{A}$, and a given number of sources, $k$, is NP-hard. \cite{tillmann2014}
However, for a fixed $k$, one can Monte Carlo over source positions and compute a Statistical Restricted Isometry Property (StRIP). \cite{calderbank2010} 
By examining the probability distribution function for the RIP parameter and evaluating it for a $95\%$ confidence interval, we can obtain a StRIP for the set of masks shown in Fig.~\ref{fig3} of $\delta^k_{95} = 0.79$.
Thus, in the statistical sense, the $\mathbf{A}$ matrix given in Eq. \eqref{eq01} satisfies $\delta_k < 1$ for our choice of masks.

To experimentally test our predicted results, a waveguide was constructed out of acrylic sheets and sealed on all sides with RTV silicone. 
The waveguide geometry is illustrated in Fig.~\ref{fig2}.  
Using a spacer layer, the waveguide height, $h=9~\mathrm{mm}$, was selected to be less than a half-wavelength at the design frequency of 10~kHz ($\lambda\sim 34.3~\mathrm{mm}$). 
This maintained a transverse planewave mode in the waveguide, simulating an infinite 2D system. 
\begin{figure}[tb]
	\begin{flushleft}
		\includegraphics[height=0.75\textheight]{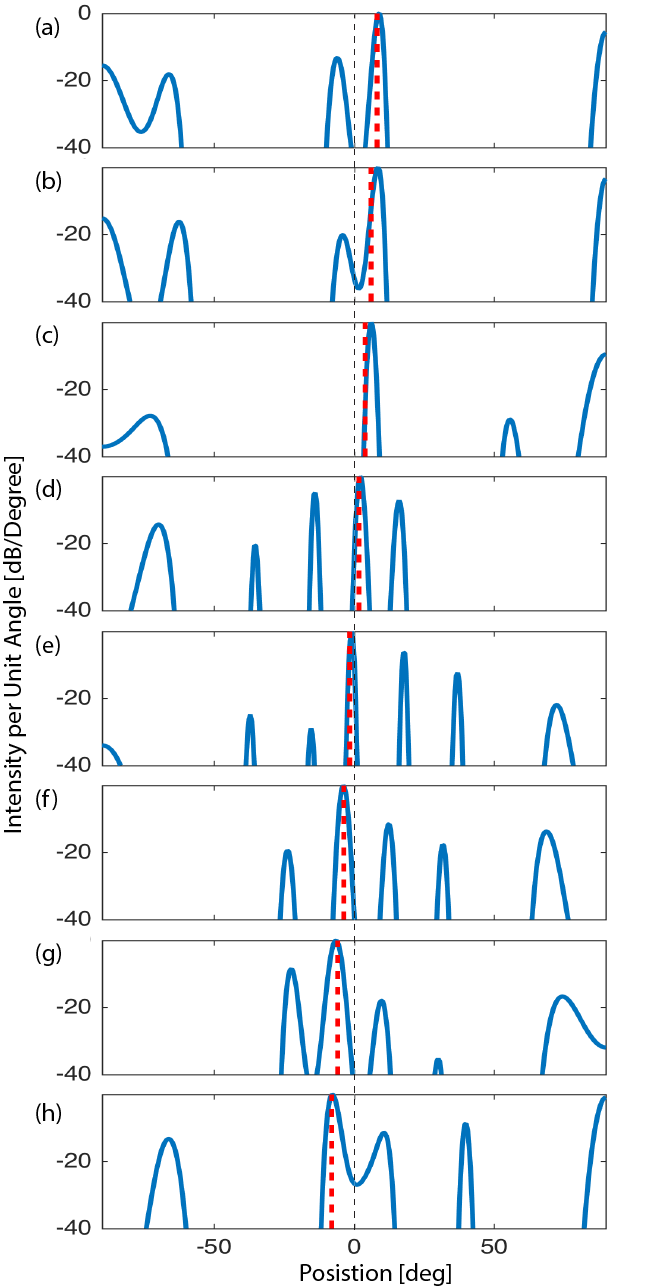}
		\caption{Retrieved $\hat{\mathbf{x}}$, the direction depended power vector, for the eight source positions (a-h) $8^\circ$,$6^\circ$,$4^\circ$,$2^\circ$,$-2^\circ$,$-4^\circ$,$-6^\circ$,$-8^\circ$. With (red, dashed) measured position indicated.}
		\label{fig4}
	\end{flushleft}
\end{figure}
Sound absorbing foam was lined along the waveguide interior to minimize edge reflections and further approximate an infinite 2D system.

An open slot was left at the waveguide center, to allow the insertion of laser cut blocking screens.
The masks are designed to have 16 unit cells each having a width of $2a=\lambda /4$ and the binary screen patterns used are those highlighted in Fig.~\ref{fig3} (a).
Due to the masks consisting of multiple adjacent open unit cells, diffraction effects can be observed in the measured pressure field as observed in Fig.~\ref{fig3} (b).
These diffraction effects can be accounted for using the model described by Eq.~(\ref{eq03}).
 Qualitatively, the model shows agreement with experimental data as depicted by comparing Fig.~\ref{fig3} (c) with Fig.~\ref{fig3} (b).

The indicated receiver in Fig.~\ref{fig2} was a Br{\"u}el and Kj\ae r 4939-A-011 microphone placed on the center axis of the waveguide, 530mm away from the aperture screen location. 
This receiving microphone was amplified by a Br{\"u}el and Kj\ae r 2690-0S4 pre-amplifier set to 1~V/Pa.

The source, a Dayton Audio 32A-8 speaker, was mounted 530~mm away from the aperture screen, on a plate with perpendicular translating motion, ruled in 1mm increments. 
The source speaker was driven by an Agilent 33220A function generator, using a programmed 1~ms duration 10~kHz center frequency pulse, windowed with a Hanning envelope function.
The source was placed at seven distinct locations: $\pm$15mm, $\pm$35mm, $\pm$55mm and $\pm$75 above (+) and below (-) the central wave guide longitudinal axis, to create the respective set of 8 azimuthal bearings, $\pm2^\circ$,$\pm4^\circ$,$\pm6^\circ$,$\pm8^\circ$. 

The collected time series data were digitized using a National Instrument cRio NI-9223 analog-to-digital converter sampling at 1MHz. 
The collection was syncronized to the function generator output to allow sample-to-sample averaging and to establish a constant time reference for measuring the full acoustic field plots. 
The collected waveforms were, for each of the ten apertures, time windowed to minimize  edge reflection noise, and 512 time series samples were averaged together for each aperture-type/source-location combination to maximize the signal to noise ratio.

The full field plots in Fig.~\ref{fig3} were taken for two different source positions for all 10 masks to  analyze error as a function of receiver position.
The back of the waveguide was removed and the receiver microphone was mounted on to a computer controlled x-y stage (Velmex VMX) with an extension rod.
Time series waveforms were collected throughout the waveguide transmission domain, and processed to retrieve the instantaneous acoustic amplitude and phase at equally spaced points ($\Delta x = \Delta y = 5~\mathrm{mm}$). 
For the full field analysis, the amplitude and phase of the acoustic wave was extracted from the collected time series data with a fast Fourier transform of a time windowed, linear frequency modulated pulse. 
Note that the masks are centered at 0 mm along the Y axis and 590 mm along the Z axis.


The retrieved $\ell_1$ normalization results following Eq. \eqref{eq09} for all tested azimuths are shown in Fig.~\ref{fig4}. 
A total of only four out of the ten highlighted masks, were needed to successfully invert for the eight different source positions. 
The masks used are outlined in red in Fig.~\ref{fig3}.
The Fig.~\ref{fig4} plots show $\mathbf{\hat{x}}(\theta)$, the retrieved power, in decibels, as a function of angular position, $\theta$. 
For all measured source positions, the maximal points in Fig.~\ref{fig4}, correspond to the known source positions (the red dashed line). 
This reflects our successful determination of source position with a highly under-sampled set of possible k-vectors. 

The additional peaks that appear in the inversions are due to grating lobes (or grating orders) caused by the high degree of periodic order inherent to the DCT basis chosen to construct the masks.
For example, in Fig.~\ref{fig3} (b), grating lobes are observed due to the large separation between openings in the middle of the mask.
These grating lobes result in spurious peaks in the inversion output and may obscure the source being localized as observed in Fig.~\ref{fig4} (h).
Here, a grating lobe caused a spurious peak near positive endfire having roughly the same magnitude as the peak corresponding to the true source direction.

\begin{figure}[h]
	\begin{center}
		\includegraphics[width=\columnwidth]{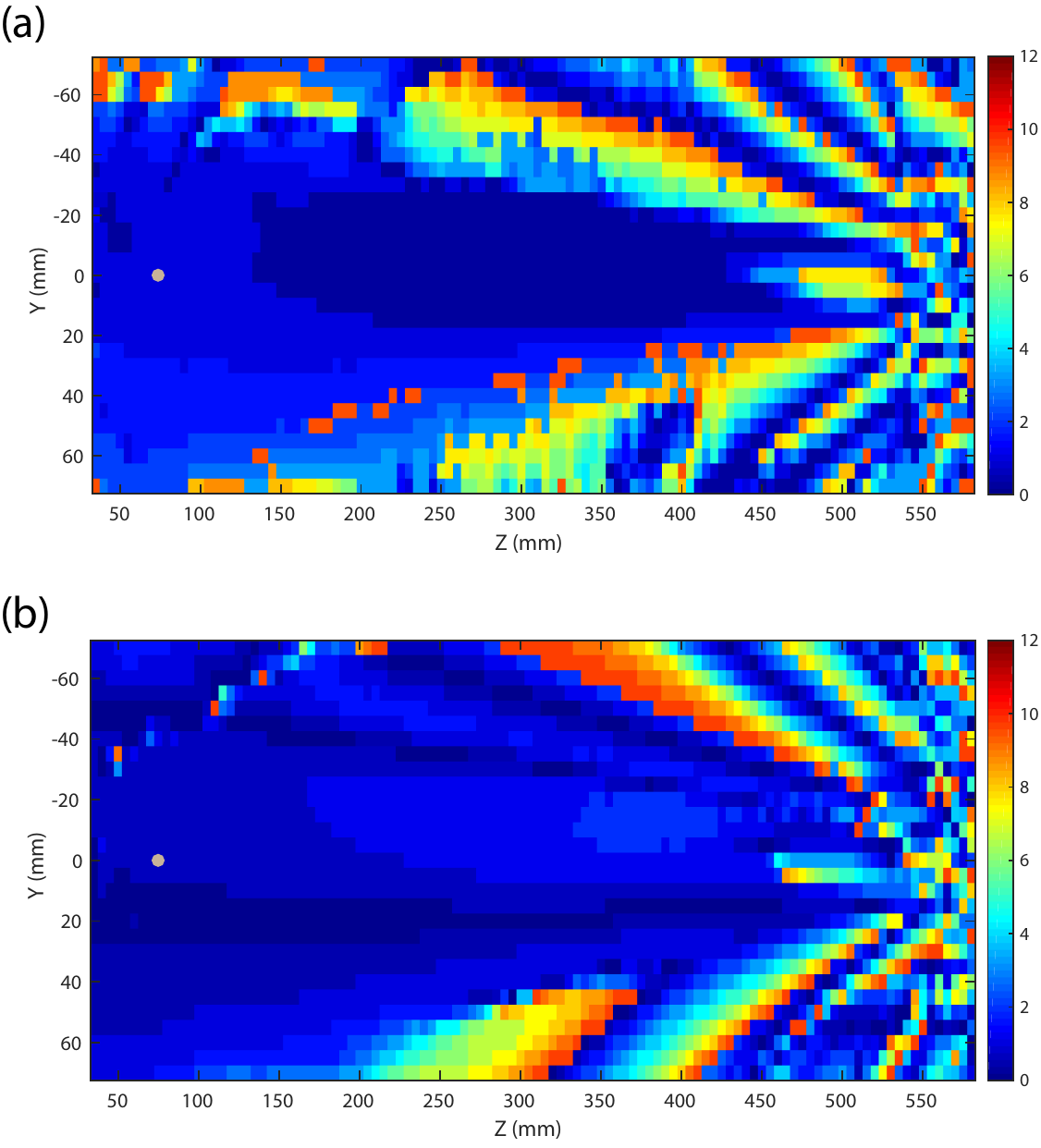}
		\caption{Target localization error surfaces computed over a grid of receiver positions within the waveguide for a source position of (a) $2^\circ$ and (b) $-6^\circ$. The position and size of the single pixel receiver is indicated with the gray dot.}
		\label{fig5}
	\end{center}
\end{figure}

In our experimental design, the grating lobes are a result of lengthening the aperture to obtain a higher resolution output. 
At the receiver, the grating lobes only appear at angles outside the physical boundary of the waveguide.
Therefore, when determining a source position, we only need to consider the maximum amplitude between $\pm 8^{\circ}$.
To better illustrate the localization performance, an error surface is shown in Fig.~\ref{fig5}.
The surface represents the localization error at different single point receiver locations within the waveugide.
The error is computed by taking the absolute value of the difference between the true source bearing and the bearing corresponding to the maximum amplitude between $\pm 8^\circ$ for a source positions of $2^\circ$ and $-6^\circ$.
The figure illustrates the effect of the grating lobes and nearfield diffraction on localization performance. 
At distances very close to the mask, there is mismatch in the diffraction model used in Eqs.~(\ref{eq03})--(\ref{eq08}).
The structured peaks emanating radially from the mask in the error surface illustrate the ambiguity caused by grating lobes from the masks.
However, localization error is minimal at positions sufficiently far from the mask and near the center of the waveguide (to minimize imperfect side wall absorption).

In this work we have presented our application of compressive imaging to build a single pixel acoustic camera. 
We have used this technique to determine the azimuthal position of an acoustic source, using an omni-directional receiver and a set of analog apertures in a simplified 2D waveguide. 
The experimental results have shown an under-sampled set of four orthogonal aperture screens allowed the determination of an acoustic source position with an accuracy of $\pm 1^\circ$ for a large portion of the acoustic waveguide. 
This research lays the ground work for building more complex single pixel imaging systems for sampling sparse acoustic targets. 
[This work was supported by ONR.]


%

\end{document}